\documentstyle[12pt,epsf]{article} 
\newcommand{\beq}{\begin{equation}}
\newcommand{\eeq}{\end{equation}}
\newcommand{\beqa}{\begin{eqnarray}}
\newcommand{\eeqa}{\end{eqnarray}}

\newcommand{\krig}[1]{\stackrel{\circ}{#1}}

\begin{document}

\pagestyle{empty}

\begin{center}

\hfill KFA-IKP(TH)-1997-03

\smallskip

%{\large { \bf  RECENT DEVELOPMENTS IN HEAVY BARYON CHIRAL \\[0.15cm]
%PERTURBATION THEORY: SELECTED TOPICS}}
{\large { \bf  Recent developments in heavy baryon chiral \\[0.15cm]
perturbation theory: Selected topics}}

\vspace{0.3cm}
                      
Ulf--G. Mei{\ss}ner\\        
{\it FZ J\"ulich, IKP (Theorie), D--52425 J\"ulich, Germany}

\vspace{0.2cm}

\end{center}

\baselineskip 10pt

\noindent I review recent results in baryon chiral perturbation
theory, in particular related to pion--nucleon scattering and
first systematic attempts to go beyond next-to-leading order in
the case of  three flavors. New insight into the chiral expansion
of the baryon masses and magnetic moments is presented.

\vspace{0.3cm}

\baselineskip 12pt

\noindent {\bf 1 $\quad$ The two--flavor case: LECs and $\pi N$ scattering}
 
\vspace{0.3cm}

\noindent Chiral perturbation theory with nucleons is by now in a fairly
mature status. It exploits the chiral symmetry properties of QCD and
makes use of methods borrowed from heavy quark effective field theory to
consistently deal with the spin--1/2 particles. A consistent power counting
in a triple expansion of external momenta, quark masses and inverse powers
of the nucleon mass (collectively denoted by the small parameter $p$)
 allows to investigate many single--nucleon processes 
as detailed in the recent review [1]. In case of two or more nucleons,
there is still active debate about how to formulate the chiral expansion
(of the potential, of the inverse amplitude, of the interaction kernel,
 $\ldots$). Although highly interesting, for the sake of brevity I will 
not further address these topics here. Instead, I will briefly high--light
two novel calculations related to the determination of the dimension two
low--energy constants and to pion--nucleon scattering. The effective 
pion--nucleon Lagrangian consists of a string of terms with increasing 
dimension. At second order, it contains seven a priori unknown coupling 
constants, the so--called low--energy constants (LECs). While these have
been determined before [1], these determinations involved some quantities 
in which large cancellations appear inducing some sizeable uncertainty.
In ref.[2], the four LECs related to pion--nucleon scattering and 
isoscalar--scalar external sources (as measured e.g. in the $\sigma$--term) 
were fixed from a set of nine observables which to one--loop order $p^3$
are given entirely by tree and loop diagrams with insertions from the 
dimension one and two parts of the effective Lagrangian. The fifth LEC is
only non--vanishing in case of unequal light quark masses and can thus be
estimated from the strong contribution to the neutron--proton mass difference.
The other two LECs are given by the anomalous magnetic moments of the proton
and the neutron. In that paper, it was also shown that the numerical values
of these LECs can indeed be understood from resonance exchange, however, in
some cases there is sizeable uncertainty related to certain $\Delta$ couplings.
Most interesting is the finding that the LEC $c_1$ reveals the strong pionic
correlations coupled to nucleons well known from phenomenological models of the
nucleon--nucleon force. 
%%% here table 1
%%%
\smallskip

\begin{tabular}{|c|c|c|r|c|}
\hline
\hline
       & ocuurs in & determined from & value & res. exch. \\
\hline
$c_1'$ & $m_N,\sigma_{\pi N}$, $\gamma N \to \gamma N$ 
              & phen. + res.exch. & $-1.7\pm 0.2 $ & $-1.7^*$ \\
$c_2'$ & $\pi N \to \pi (\pi) N$, $\gamma N \to \gamma N$     
              & phen. + res.exch. & $6.3\pm 0.4 $ & $3.8 \ldots 7.5$ \\
$c_3'$ & $\pi N \to \pi (\pi) N$, $\gamma N \to \gamma N$     
              & phen. + res.exch. & $-9.9\pm 0.5 $ & $-8.4 \ldots -10.0$ \\
$c_4'$ & $\pi N \to \pi (\pi) N$ & phen. + res.exch. 
                                             & $6.8 $ & $5.8 \ldots 6.9$ \\
$c_5'$ & $(m_n-m_p)^{\rm strong}, \pi^0 N \to \pi^0 N $  
                                 & phenomenology & $-0.17 $ & --- \\
$c_6$ & $\kappa_p$, $\kappa_n$   & phen. + res.exch. & $5.8 $ & $6.1$ \\
$c_7$ & $\kappa_p$, $\kappa_n$   & phen. + res.exch. & $-3.0 $ & $-3.1$ \\
\hline
\hline
\end{tabular}\smallskip

\noindent Table~1: Values of the dimension two LECs $c_i ' = 2mc_i$
$(i=1,\ldots,5)$ as determined in [2]. Also given are the ranges based
on resonance saturation. The uncertainties and ranges are discussed
in detail in that reference. The $^*$ denotes an input quantity.

\noindent The one-loop contribution to the $\pi N$-scattering amplitude 
to order $p^3$ has first been worked out by Moj\v zi\v s [6]. Here, I 
follow ref.[2] in which certain aspects of pion--nucleon scattering have 
also been addressed. In the center-of-mass frame the $\pi
N$-scattering amplitude $\pi^a(q) + N(p) \to \pi^b(q') + N(p')$ takes the
following form: 
\begin{equation} 
T^{ba}_{\pi N} = \delta^{ba} \Big[ g^+(\omega,t)+ i \vec
\sigma \cdot(\vec q\,'\times \vec q\,) \, h^+(\omega,t) \Big] +i \epsilon^{bac}
\tau^c \Big[ g^-(\omega,t)+ i \vec \sigma \cdot(\vec q\,'\times \vec q\,) \,
h^-(\omega,t) \Big] 
\end{equation}
with $\omega = v\cdot q = v\cdot q\,'$ the pion cms energy and $t=(q-q\,')^2$ 
the invariant momentum transfer squared. $g^\pm(\omega,t)$ refers to the
isoscalar/isovector non-spin-flip amplitude and $h^\pm(\omega,t)$ to the
isoscalar/isovector spin-flip amplitude. After renormalization of the pion
decay constant $F_\pi$ and the pion-nucleon coupling constant $g_{\pi N}$, one
can give the one-loop contributions to the cms amplitudes
$g^\pm(\omega,t)$ and $h^\pm(\omega,t)$ at order $p^3$ in closed form, 
see ref.[2].
The $t$-dependences of the loop-amplitudes $g^\pm(\omega,t)_{\rm loop}$ and
$h^\pm(\omega,t)_{\rm loop}$ show an interesting structure, if one discards
terms proportional to $g_A^4$. The $t$-dependence of
$h^+(\omega,t)_{\rm loop}$ is then given by $(2t-M_\pi^2)/(3M_\pi^2F_\pi^2)\,
\sigma(t)_{\rm loop}$, with $\sigma(t)$ the nucleon scalar form
factor. Furthermore, the $t$-dependence of $g^-(\omega,t)_{\rm loop}$ becomes
equal to $\omega/(2F_\pi^2)\,G_E^V(t)_{\rm loop}$, with $G_E^V(t)$ the nucleon
isovector electric form factor (normalized to unity). Finally, $h^-(\omega,t)_{
\rm loop}$ has the same $t$-dependence as $-1/(4mF_\pi^2)\,  G_M^V(t)_{\rm
loop}$, with $G_M^V(t)$ the  nucleon isovector magnetic form factor. The
one-loop calculation of these nucleon form factors can be found in [7]. 
In table 2, I show the predictions for the remaining S, P, D and
F-wave threshold parameters which were not used in the fit to determine the
LECs. In some cases, contributions from the dimension three Lagrangian appear.
The corresponding LECs have been estimated using resonance exchange. In 
particular, the 10\% difference in the P--wave scattering volumina $P_1^-$
and $P_2^+$ is a clear indication of chiral loops, because
nucleon and $\Delta$ Born
terms give the same contribution to these two observables. Note also that
the eight D-- and F--wave threshold parameters to this order are free of
contributions from dimension three and thus uniquely predicted. The overall
agreement of the predictions with the existing experimental values is rather
satisfactory.

\renewcommand{\arraystretch}{1.3}

\begin{center}

\begin{tabular}{|c|c|c|c|c|c|c|}
    \hline  \hline
    Obs. & CHPT &  Order & Ref. & Exp. value & Ref. & Units \\
    \hline
$a^+$ &  $9.2\pm 0.4$ & $p^4$ & [3] & $ 8.4 \ldots 10.4$ &
[4] & $10^{-2}\, M_\pi^{-1} $ \\
$b^-$ &  $2.01$ & $p^3$ & [2] & $ 1.32 \pm 0.62$ &
[5] & $10^{-2}\, M_\pi^{-1} $ \\
   \hline
$P_1^-$ &  $-2.44\pm 0.13$ & $p^3$ & [2] & $ -2.52 \pm 0.03$ &
[5] & $ M_\pi^{-3} $ \\
$P_2^+$ &  $-2.70\pm 0.12$ & $p^3$ & [2] & $ -2.74 \pm 0.03$ &
[5] & $ M_\pi^{-3} $ \\
   \hline
$a^+_{2+}$ &  $-1.83$ & $p^3$ & [2] & $ -1.8 \pm 0.3$ &
[5] & $10^{-3}\, M_\pi^{-5} $ \\
$a^+_{2-}$ &  $2.38$ & $p^3$ & [2] & $ 2.20 \pm 0.33$ &
[5] & $10^{-3}\, M_\pi^{-5} $ \\
$a^-_{2+}$ &  $3.21$ & $p^3$ & [2] & $ 3.20 \pm 0.13$ &
[5] & $10^{-3}\, M_\pi^{-5} $ \\
$a^-_{2-}$ &  $-0.21$ & $p^3$ & [2] & $ 0.10 \pm 0.15$ &
[5] & $10^{-3}\, M_\pi^{-5} $ \\
   \hline
$a^+_{3+}$ &  $0.29$ & $p^3$ & [2] & $ 0.43 $ &
[5] & $10^{-3}\, M_\pi^{-7} $ \\
$a^+_{3-}$ &  $0.06$ & $p^3$ & [2] & $ 0.15 \pm 0.12$ &
[5] & $10^{-3}\, M_\pi^{-7} $ \\
$a^-_{3+}$ &  $-0.20$ & $p^3$ & [2] & $ -0.25 \pm 0.02$ &
[5] & $10^{-3}\, M_\pi^{-7} $ \\
$a^-_{3-}$ &  $0.06$ & $p^3$ & [2] & $ 0.10 \pm 0.02$ &
[5] & $10^{-3}\, M_\pi^{-7} $ \\
   \hline   \hline
  \end{tabular} \end{center}
\medskip

\noindent Table~2: Threshold parameters predicted by CHPT. The order
of the prediction is also given together with the experimental values.

\bigskip
\vspace{0.5cm}

\noindent {\bf 2 $\quad$ Three flavors: General aspects}
 
\vspace{0.3cm}

\noindent Heavy baryon CHPT was originally formulated for the
three--flavor case [8], however, most of the numerous calculations
performed only included the leading and sub--leading non-analytic 
corrections and some
counter terms. Furthermore, it was argued that the spin--3/2 decuplet
should be included in the effective field theory since it was found
that in many cases the large kaon loop corrections were cancelled by
diagrams with intermediate decuplet states. For an early review on these
activities, see ref.[9]. To really assess the validity of the approach, 
it is mandatory to perform calculations that include {\it all} terms at
the given order and should also be extended to order $p^4$ in the chiral 
expansion, guided by the experience from the two--flavor sector. Clearly,
one can not expect  calculations in SU(3) to be as precise as their SU(2)
counterparts (for the same order in the chiral expansion), simply because
the expansion parameter is significantly larger,
\beq    
{\rm SU(2):} \,\,\,\, {M_\pi \over 4 \pi F_\pi} \simeq 0.1 \, , \quad
{\rm SU(3):} \,\,\,\, {M_K \over 4 \pi F_K} \simeq 0.4 \,\,.
\eeq 
Similarly, the thresholds in scattering or production processes are
much higher in energy. For example, producing neutral pions off protons
by real photons has a threshold energy of about $E_\gamma = $145~MeV, whereas 
the production of neutral kaons only starts at $E_\gamma \simeq $900~MeV.
In what follows, I will present the results of some calculations which try
to systematically explore the applicability and limitations of the 
three--flavor approach. Before discussing these specific examples, let
me turn to some more theoretical aspects, i.e. the problem that to one
loop in the chiral expansion divergences appear. 
The divergence structure of the one--loop generating functional to order
$p^3$ has been worked out in ref.[10]. It extends previous works by Ecker
and Moj\v zi\v s [11,12] for the pion--nucleon Lagrangian to the SU(3) 
case. While the method outlined in [11] can also be used in SU(3), the fact 
that the baryons are in the adjoint representation of SU(3) whereas the
nucleons are in the fundamental representation of SU(2), complicates the
calculations considerably. In fig.~1 the various contributions to the 
one--loop generating functional together with the tree level generating 
functional at order $\hbar$ are shown. The solid (dashed) double lines 
represent the baryon (meson) propagator in the presence of external fields.
Only if one ensures that the field definitions underlying the mesonic
and the baryon-meson Lagrangian match, the divergences are entirely given
by the irreducible self--energy ($\Sigma_1$) and the tadpole ($\Sigma_2$)
graphs. 

\medskip

\hskip 1.5in
\epsfxsize=3in
\epsffile{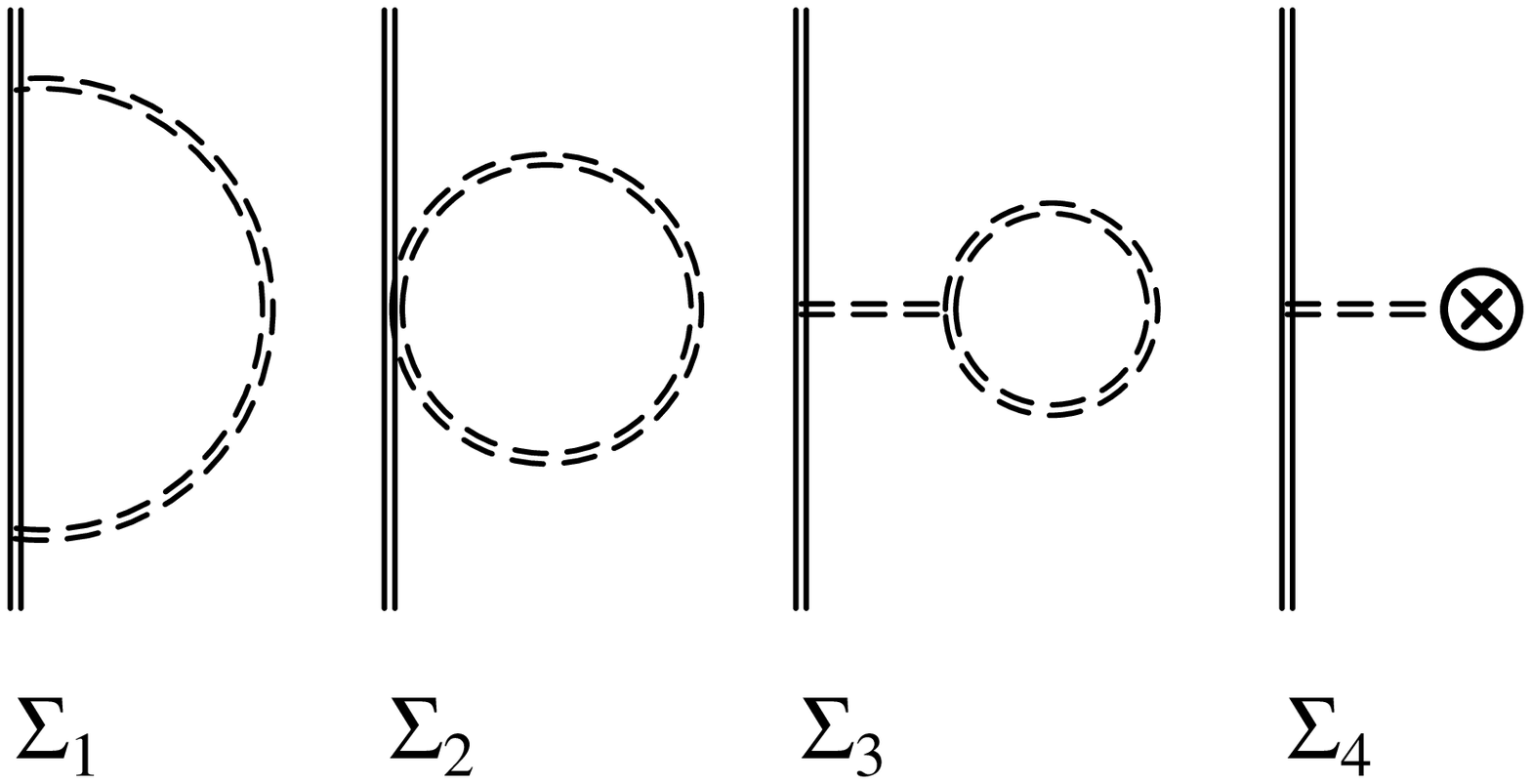}

\centerline{Fig.~1: Contributions to the one--loop generating functional 
at order $\hbar$.}

\medskip

\noindent The explicit calculations to extract the divergences 
from $\Sigma_{1,2}$ are given in [10]. The generating functional can be 
renormalized by introducing the following counterterm Lagrangian
\beq
{\cal L}_{MB}^{(3) \,{\rm ct}} = \frac{1}{(4\pi F_\pi)^2} \sum_{i=1}^{102}
\, d_i \, \bar{H}^{ab}_v (x) \, O_i^{bc}(x) \,{H}^{ca}_v (x) \,\,\, ,
\eeq
with $'a,b,c'$ SU(3) indices, ${H}^{ab}_v (x)$ denotes the velocity
eigenstate with eigenalue $+1$ of the heavy baryon field and the field
monomials $O_i^{bc}(x)$ are of order $p^3$. The dimensionless LECs
$d_i$ are decomposed as
\beq
d_i = d_i^r (\mu) + (4\pi)^2 \, \beta_i \, L(\mu) \,\, ,
\eeq
with
\beq
L(\mu) =\frac{\mu^{d-4}}{(4\pi)^2} \biggl\{ \frac{1}{d-4} - \frac{1}{2}
\bigl[ \log(4\pi) + 1 - \gamma \bigr]\biggr\} \,\, .
\eeq
Here, $\mu$ is the scale of dimensional regularization, $\gamma$ the 
Euler--Mascheroni constant and the $\beta_i$
are dimensionless functions of the axial couplings $F$ and $D$ that
cancel the divergences of the one--loop functional. They are tabulated
in [10] together with the $O_i^{bc}(x)$. These 102 terms constitute a 
complete set for the renormalization with off--shell baryons.  
As long as one is only interested in Greens functions
with on--shell baryons, the number of terms can be reduced considerably
making use of the baryon equations of motion. Also, many of these terms 
involve processes
with three or more mesons. So for calculations of kaon--nucleon scattering
or kaon photoproduction off nucleons, many of these terms do not contribute
(or only start to contribute at higher orders). An example will be given 
below. At present, only few of the finite $d_i^r (\mu)$ have been determined.
There are two main directions to extend these investiagtions. First, a 
systematic effort to pin down as many LECs as possible is needed 
and, second, the divergences at order $p^4$ should be extracted. Work 
along these lines is underway.

\vspace{0.5cm}

\noindent {\bf 3 $\quad$ The scalar sector: Baryon masses and $\sigma$--terms}
 
\vspace{0.3cm}

\noindent The scalar sector of baryon CHPT is particularly interesting since
it is sensitive to scalar--isoscalar operators and thus directly to the
symmetry breaking of QCD. This is most obvious for the pion-- and 
kaon--nucleon $\sigma$--terms, which measure the strength of the scalar
quark condensates $\bar q q$ in the proton. Here, $q$ is a generic symbol for 
any one of the light quarks $u$, $d$ and $s$. Furthermore, the quark mass
expansion of the baryon masses allows to gives bounds on the ratios of the
light quark masses [13]. The most general effective Lagrangian to fourth 
order in the small parameter $p$  necessary to investigate the scalar sector 
consists of seven dimension two and seven dimension four terms
with LECs plus some additional dimension two terms with fixed 
coefficients $\sim 1/m$. The dimension two terms with LECs fall into 
two classes, one related to symmetry breaking and the other are 
double--derivative meson-baryon vertices. The LECs related to the latter 
ones can be estimated with some confidence from resonance exchange. 
A method to estimate the symmetry breakers will be discussed below. 
The analysis of the octet baryon masses in the framework of chiral
perturbation theory already has a long history, see e.g. [14]. However, only
recently the results of a  calculation including  all terms of second 
order in the light quark masses, ${\cal O}(m_q^2)$, were presented [15]. 
The calculations were performed  in the isospin limit $m_u = m_d$ and the 
electromagnetic corrections were neglected. Previous investigations considered
mostly the so--called computable corrections of order $m_q^2$ or included 
some of the finite terms at this order. This, however, contradicts the spirit 
of CHPT in that all terms at a given order have to be retained. 
The quark mass expansion of the octet baryon masses takes the form
\begin{equation}
m = \, \, \krig{m} + \sum_q \, B_q \, m_q + \sum_q \, C_q \, m_q^{3/2} + 
\sum_q \, D_q \, m_q^2  + \ldots
\label{massform}
\end{equation}
modulo logs. Here, $\krig{m}$ is the octet mass in the chiral limit of
vanishing quark masses and the coefficients $B_q, C_q, D_q$ are 
state--dependent. Furthermore, they include contributions proportional
to some LECs which appear beyond leading order in the effective Lagrangian.
In contrast to the ${\cal O}(p^3)$ calculation, which gives the leading
non-analytic terms $\sim m_q^{3/2}$, the order $p^4$ one is no longer finite
and thus needs renormalization. Intimately connected to the baryon masses
are the $\sigma$--terms, which are defined in a mass--independent 
renormalization scheme via, 
\begin{eqnarray}
\label{defsigma}
&&\sigma_{\pi N} (t)  =  \hat m \, \langle p' \, 
| \bar u u + \bar d d| \, p \rangle
\, \, \, , \sigma_{KN}^{(1)} (t) = \frac{1}{2}(\hat m + m_s) \, 
\langle p' \, | \bar u u + \bar s s| \, p \rangle \, \, \, ,  \nonumber \\
&&\sigma_{KN}^{(2)} (t)  = \frac{1}{2}(\hat m + m_s) \, 
\langle p' \, | -\bar u u + 2\bar d d + \bar s s| \, p \rangle \, \, \, , 
\end{eqnarray}
with $|p\rangle$ a proton state with four--momentum $p$ and $t 
= (p'-p)^2$ the 
invariant momentum transfer squared. A relation between $\sigma_{\pi N} (0)$
and the nucleon mass is provided by  the Feynman--Hellmann theorem, 
$\hat{m}(\partial m_N / \partial \hat{m}) = \sigma_{\pi N} (0)$, with 
$\hat m$ the average light quark mass. Furthermore, the strangeness fraction 
$y$ and $\hat \sigma$ are defined via
\begin{equation}
y = \frac{ 2 \, \langle p| \bar  s s|p \rangle}
{\langle p|\bar u u + \bar d d |p \rangle} =
\frac{M_\pi^2}{\sigma_{\pi N} (0)} \biggl( M_K^2 -
\frac{ M_\pi^2 }{2}  \biggr)^{-1}   \, 
 m_s \, \frac{\partial \,m_N}{\partial \,m_s} \equiv 1 -
\frac{\hat \sigma}{\sigma_{\pi N} (0)} \,\, .
\label{defy}
\end{equation}
Let me turn to the calculations presented in [15]. As stated before, there
are ten LECs related to symmetry breaking. Since there do not exist enough 
data to fix all these, they were estimated  by means of resonance exchange. 
To deal with such scalar-isoscalar operators, the standard resonance 
saturation picture based on tree graphs was  extended to include loop diagrams.
This is also done in calculations of the deviations from Dashen's theorem,
where one considers loops with photons and heavy (axial)vector mesons [16]. In
contrast to the two-flavor case, the scalar mesons in SU(3) can not explain 
the strength of the symmetry breakers because these mesons
 are not effective degrees
of freedom parametrizing strong pionic/kaonic correlations. To be precise,
the dimension two symmetry breakers can be estimated by performing a 
best fit ot the baryon masses based on a ${\cal O}(p^3)$ calculation [17].
For scalar couplings of ``natural'' size, these values can not  even
be reproduced wihin one order of magnitude.
One way to solve this problem, although it has its own conceptual
difficulties, is to consider besides standard tree graphs with scalar meson 
exchange also Goldstone boson loops with intermediate baryon resonances 
(spin--3/2 decuplet and the spin--1/2 (Roper) octet)
for the scalar--isoscalar LECs. In [15] a consistent scheme to implement 
resonance exchange under such circumstances was developed. In particular, it
avoids double--counting and abids to the strictures from analyticity. Within 
the one--loop approximation and to leading order in the resonance masses, the 
analytic pieces of the pertinent graphs are still divergent, i.e.
one is left with three a priori undetermined renormalization constants
($\beta_\Delta$, $\delta_\Delta$ and $\beta_R$). These have to be
determined together with the finite scalar meson--baryon couplings $F_S$ 
and $D_S$ and the octet mass in the chiral limit. Using the baryon masses and
the value of $\sigma_{\pi N} (0)$ as input, one can determine all LECs
in terms of one parameter, $\beta_R$. This parameter can be shown to be 
bounded and the  observables are insensitive to variations of it within 
its allowed range. Furthermore, it was also demonstrated
that the effects of two (and higher) loop diagrams can almost entirely
be absorbed in a redefinition of the one loop renormalization parameters. 
Within this scheme, one finds for the octet baryon mass in the chiral limit
$\krig{m} = 770\pm 110\, {\rm MeV}$. The quark mass expansion of the baryon 
masses,  in the notation of Eq.(\ref{massform}), reads
\begin{eqnarray}
&& m_N  = \,  \krig{m}  \, ( 1 + 0.34 - 0.35 + 0.24 \, ) \, \, ,
\nonumber \\
&& m_\Lambda  = \,  \krig{m} \, ( 1 + 0.69 - 0.77 + 0.54 \, ) \, \, ,
\nonumber \\
&& m_\Sigma  = \,  \krig{m} \, ( 1 + 0.81 - 0.70 + 0.44 \, ) \, \, ,
\nonumber \\
&& m_\Xi  = \,  \krig{m} \, ( 1 + 1.10 - 1.16 + 0.78 \, ) \, \, .   
\label{mexpand}
\end{eqnarray}
One observes that there are large cancellations between the second order
and the leading non--analytic terms of order $p^3$, a well--known effect.
The fourth order contribution to the nucleon mass is fairly small, whereas it
is sizeable for the $\Lambda$, the $\Sigma$ and the $\Xi$. This is
partly due to the small value of $\krig{m}$, e.g. for the $\Xi$ the
leading term in the quark mass expansion gives only about 55\% of the
physical mass and the second and third order terms cancel almost completely.
From the chiral expansions exhibited in Eq.(\ref{mexpand}) one can not    
yet draw a final conclusion about the rate of convergence in the
three--flavor sector of baryon CHPT. Certainly, the breakdown of CHPT
claimed in [13] is not observed. On the other hand, the
conjecture [18]  that only the leading non--analytic corrections (LNAC)
$\sim m_q^{3/2}$ are large and that further terms like the ones $\sim m_q^2$
are moderately small, of the order of 100 MeV, is not supported. 
The chiral expansion of the $\pi N$ $\sigma$--term shows a moderate 
convergence, i.e. the terms of increasing order become successively smaller,
\begin{equation}
\sigma_{\pi N} (0) = 58.3 \, ( 1 - 0.56  + 0.33) \, \, \, {\rm MeV} 
= 45 \, \, {\rm MeV}     \, \, .
\label{signo}
\end{equation}
Still, the $p^4$ contribution is important.  For the strangeness fraction  
$y$ and $\hat \sigma$, one finds
\begin{equation}
y = 0.21 \pm 0.20 \, \, , \, \,
\hat \sigma = 36 \pm 7 \, \, {\rm MeV} \, \, .
\label{valunc}
\end{equation}
The value for $y$ is within the band deduced in [19], $y =
0.15 \pm 0.10$ and the value for $\hat \sigma$ compares favourably
with Gasser's estimate, $\hat \sigma = 33 \pm 5\,$MeV [13]. Further 
results concerning the kaon--nucleon $\sigma$--terms and some two--loop
corrections to the nucleon mass can be found in [15]. Finally, two more
comments concerning the difference of the pion--nucleon $\sigma$--term 
at $t=0$ and at the Cheng--Dashen point are in order. First, in [20] 
it was shown that the remainder $\Delta_R$ not fixed by chiral symmetry, 
i.e. the
difference between the on--shell $\pi N$ scattering amplitude $\bar{D}^+
(0,2M_\pi^2)$ and the scalar form factor $\sigma_{\pi N} (2M_\pi^2)$,
contains no chiral logarithms and vanishes simply as $M_\pi^4$ in the
chiral limit. In addition, an upper limit was reported, 
$\Delta_R \le 2\,$MeV. Second, in [17] it was shown that a one--loop
diagram with an intermediate $\Delta (1232)$ allows to explain the numerical
value of the scalar form factor (ff). The leading $p^3$ graph with nucleon
intermediate states gives only $7.4\,$MeV, i.e. half of the empirical value
[19]. The $\Delta$-contribution, which formally starts at order $p^4$, adds
another $7.5\,$MeV. However, it was already stressed in [17] that
 the spectral function
Im~$\sigma_{\pi N} (t)/t^2$ is much less peaked around $t = 4M_\pi^2$
than the empirical one given in [19], see also fig.~2. The 
$\Delta$-contribution enhances the tail of the spectral function  
at larger $t$,
in contrast to the strong pionic correlations (higher loop effects),
which tend to enhance the spectral function close to threshold. Furthermore,
the SU(3) calculation of ref.[15]  indicates fairly sizeable strangeness
effects in this quantity. More detailed higher order calculations are
necessary to clarify this issue.

\medskip

\hskip 1.1in
\epsfxsize=3.5in
\epsffile{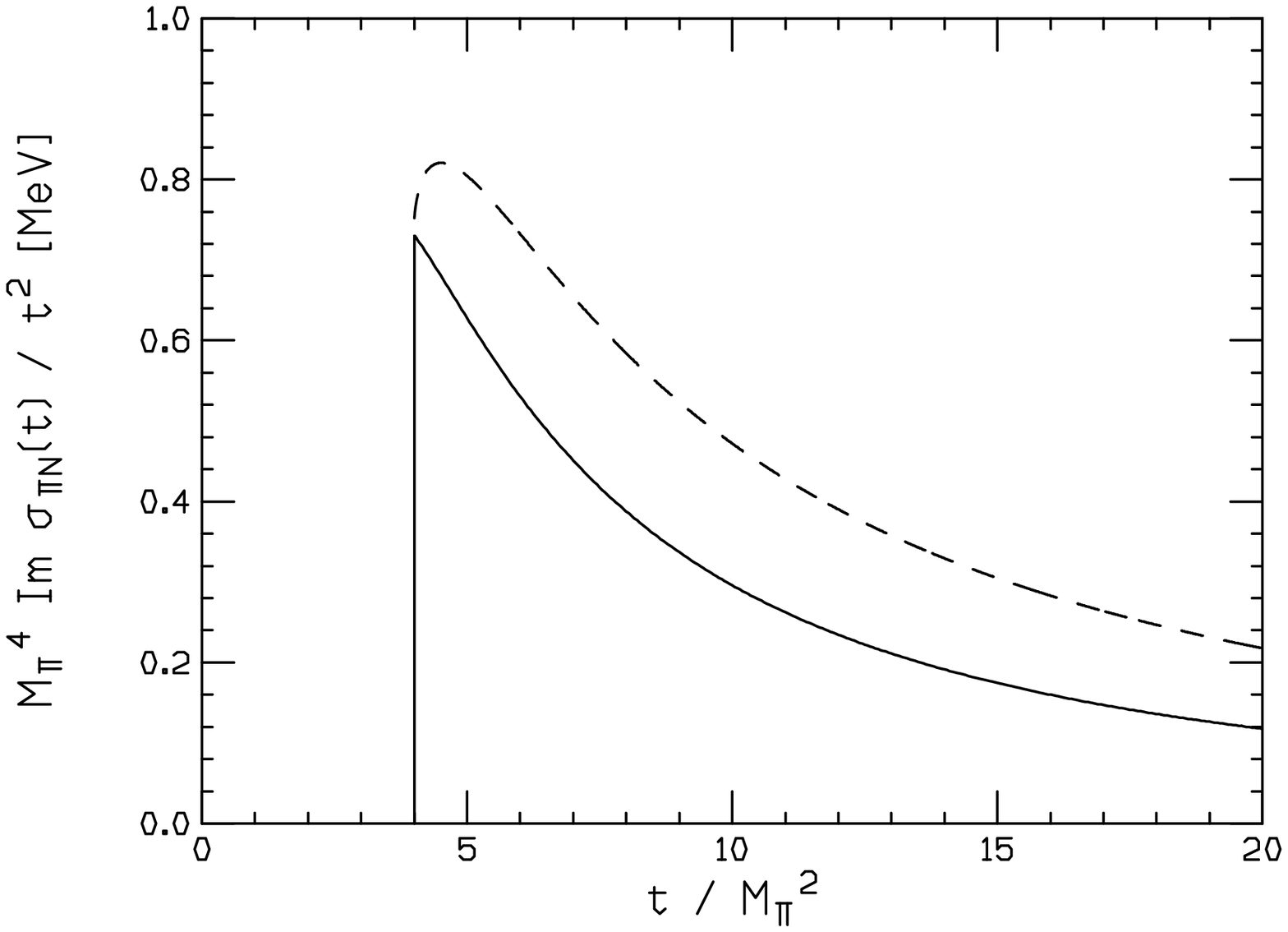}

\centerline{Fig.~2: Spectral function of the scalar ff. 
Solid line: $N$, dashed  line: $N+\Delta$ contribution.}

\bigskip

\vfill \eject

\noindent {\bf 4 $\quad$ The tale of the magnetic moments}

\vspace{0.3cm}

\noindent The magnetic moments of the octet baryons  have been measured with
high precision over the last decade. On the theoretical side, SU(3)
flavor symmetry was first used by Coleman and Glashow [21] to  
predict seven relations between the eight moments of the $p$, $n$,$\Lambda$,
$\Sigma^\pm$, $\Sigma^0$, $\Xi^-$, $\Xi^0$ and the $\mu_{\Lambda \Sigma^0}$ 
transition moment in terms of two parameters.  One of these relations is
in fact a consequence of isospin symmetry alone. In modern language, 
this was a tree level calculation with the lowest order effective
chiral meson--baryon Lagrangian of dimension two, see fig.~3a,
\begin{equation} \label{LMB2}
{\cal L}_{MB}^{(2)} = 
-\frac{i}{4m} \, b_6^F \, \langle \bar{B} [S^\mu,
S^\nu][ F_{\mu \nu}^+, B] \rangle   
-\frac{i}{4m} \,b_6^D \, \langle \bar{B} [S^\mu,
S^\nu]\{ F_{\mu \nu}^+, B\} \rangle   
\end{equation}
with $S_\mu$ the covariant spin--operator, $F_{\mu \nu}^\dagger =
-e(u^\dagger Q F_{\mu\nu}u+ uQF_{\mu\nu}u^\dagger )$ and $\langle
\ldots \rangle$ denotes the trace in flavor space. 
Here, $Q={\rm diag}(2,-1,-1)/3$ is the quark charge matrix, $u = \sqrt{U} =
\exp(i\phi/2F_\pi)$
and $F_{\mu \nu}$ the conventional photon field strength tensor.
Given the simplicity of this approach, these relations work remarkably well,
truely a benchmark success of SU(3).

\medskip

\hskip 1.7in
\epsfxsize=2.1in
\epsffile{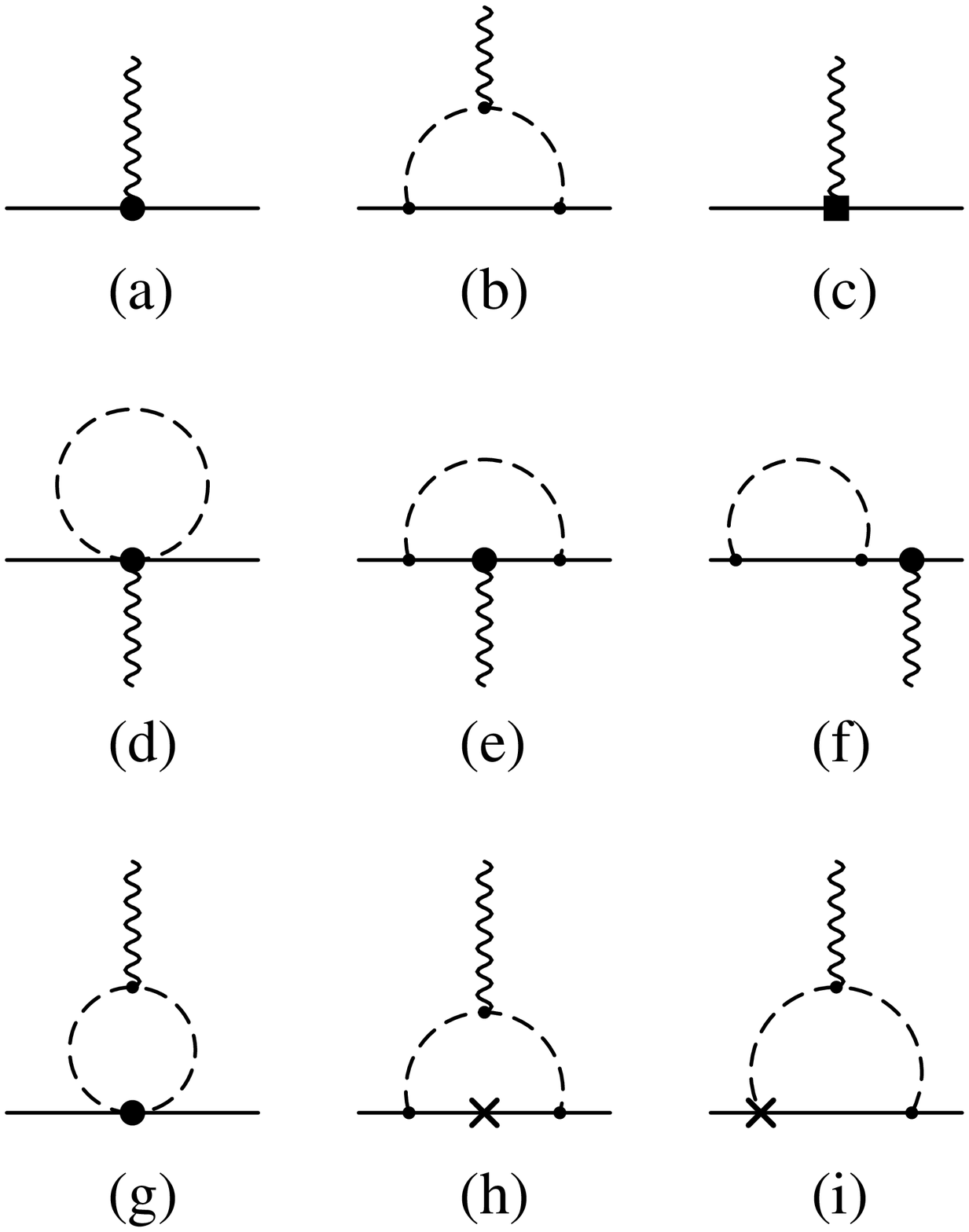}

\centerline{Fig.~3: Chiral expansion of the magnetic moments
to order $p^2$ (a), $p^3$ (b) and $p^4$ (c-i).} 

\bigskip 

\noindent The first loop corrections arise at order $p^3$ in 
the chiral counting [22], see fig~3b. They are given entirely in 
terms of the lowest order parameters from the dimension one (two)
meson--baryon (meson) Lagrangian. It was
found that these loop corrections are large for standard values of the
two axial couplings $F$ and $D$.  Caldi and Pagels [22] derived
three relations independent of these coupling constants.
These are, in fact, in good agreement with the data. However, the deviations
from the Coleman--Glashow relations get considerably worse. This fact has
some times been taken as an indication for the breakdown of SU(3) CHPT.
To draw any such conclusion, a calculation of order $p^4$ is mandatory.
This was attempted in [23], however, not all terms were accounted for.
To be precise, in that calculation the contribution from the
graphs corresponding to fig.3c-f were worked out. As pointed out in [24],
there are additional one--loop graphs at ${\cal O}{(p^4)}$, namely tadpole
graphs with double--derivative meson--baryon vertices (fig.~3g) and diagrams
with fixed $1/m$ insertions from the dimension two Lagrangian, see fig.~3h,i 
(these could
also be obtained by use of reparametrization invariance). In total, there
are seven LECs related to symmetry breaking and three related to scattering
processes (the once appearing in the graphs fig.~3g). These latter LECs
can be estimated with some accuracy from resonance exchange. The strategy
of [24] was to leave the others as free parameters and fit the magnetic
moments. One is thus able to investigate the chiral expansion of the
magnetic moments and to predict the ${\Lambda \Sigma^0}$ transition moment.
The chiral expansion of the various magnetic moments thus takes the
form
\beq 
\label{mubform}
\mu_B = \mu_B^{(2)} + \mu_B^{(3)} + \mu_B^{(4)}
= \mu_B^{(2)} \,( \, 1 + \varepsilon^{(3)} + \varepsilon^{(4)} \, ) \quad ,
\eeq
with the result (all numbers in nuclear magnetons)
\begin{eqnarray} \label{conv}
\mu_p          &=& \,\,\,\,  4.69 \, ( 1 - 0.57 + 0.16 ) =   
                                        \,\,\,\,\,  2.79 \,\, , \nonumber \\
\mu_n          &=& -2.85 \, ( 1 - 0.36 + 0.03 ) =  -1.91 \,\, , \nonumber \\
\mu_{\Sigma^+} &=& \,\,\,\,  4.69 \, ( 1 - 0.72 + 0.24 ) =  
                                          \,\,\,\,\, 2.46 \,\, , \nonumber \\
\mu_{\Sigma^-} &=& -1.83 \, ( 1 - 0.41 + 0.04 ) =  -1.16 \,\, , \nonumber \\
\mu_{\Sigma^0} &=& \,\,\,\,   1.43 \, ( 1 - 0.93 + 0.38 ) = 
                                          \,\,\,\,\, 0.65 \,\, ,           \\
\mu_{\Lambda}  &=& -1.43 \, ( 1 - 0.93 + 0.35 ) =  -0.61 \,\, , \nonumber \\
\mu_{\Xi^0}    &=& -2.85 \, ( 1 - 0.95 + 0.39 ) =  -1.25 \,\, , \nonumber \\
\mu_{\Xi^-}    &=& -1.83 \, ( 1 - 0.86 + 0.22 ) =  -0.65 \,\, , \nonumber \\
\mu_{\Lambda \Sigma^0}    
               &=&  \,\,\,\,  2.47 \, ( 1 - 0.57 + 0.18 )  = 
                                           \,\,\,\,\, 1.51 \,\, . \nonumber  
\end{eqnarray}
In all cases the ${\cal O}(p^4)$ contribution is smaller than the one 
from order $p^3$ by at least a factor of two, in most cases even by 
a factor of three. 
Like in the case of the baryon masses [15], one finds sizeable
cancellations between the leading and next--to--leading order terms
making a {\it precise} calculation of the ${\cal O}(p^4)$ terms absolutely
necessary. In fact, in all (but one) cases the contribution from the 
double--derivative terms previously omitted is the largest at order $p^4$,
one fings e.g. for the proton  $\mu_p^{(4,c)} = 1.93$, 
$\mu_p^{(4,d+e+f)} = 2.87$, $\mu_p^{(4,g)}= -4.71$ and 
$\mu_p^{(4,h+i)}= 0.71$ leading to the total of $\mu_p^{(4)}= 0.79$.
We predict the transition moment to be $\mu_{\Lambda \Sigma^0} 
= (1.51 \pm 0.01) \mu_N$ in good agreement with a recent lattice gauge theory 
result, $\mu_{\Lambda \Sigma^0} = (1.54 \pm 0.09) \mu_N$  [25]. Of
course, this is not quite the end of the story. What is certainly missing
is a deeper understanding of the numerical values of the symmetry breaking 
LECs which were used as fit parameters in [24].

\bigskip

\vspace{0.5cm}

\noindent {\bf 5 $\quad$ Kaon photoproduction}
 
\vspace{0.3cm}

\noindent Pion photo-- and electroproduction in the threshold
region has been studied intensively  over the last few years by Bernard, 
Kaiser and myself (for some recent references see [26]) with
high precision data coming from MAMI, SAL and NIKHEF. In addition,
at the electron stretcher ring ELSA (Bonn) ample kaon photoproduction data
have been taken over a wide energy range. Only a small fraction of these
data is published [27], the larger fraction is still in the process of being
analyzed. It therefore seems timely to study the reactions $\gamma p \to
\Sigma^+ K^0$, $\Lambda K^0$ and $\Sigma^0 K^+$ in the framework of CHPT. 
This has been done in an exploratory study by Steininger [28], some of the
results being published in [29]. Here, I will critically summarize the status
of these calculations.  The threshold energies for these three processes are
$1046$, $1048$ and $911\,$MeV, in order. In the threshold region, for energies
less than $100\,$MeV above the respective threshold, it is advantageous to
perform a multipole decomposition. It suffices to work with S-- and P--wave 
multipoles (the size of the D--waves has been estimated in [28]).
The transition current for the process $\gamma^* (k) +
p(p_1) \to M(q) + B(p_2)$ ($M = K^+, K^0$, $B = \Lambda, \Sigma^0, \Sigma^+$)
calculated to ${\cal O}(p^3)$ can be decomposed into Born, one--loop 
and counterterm contributions,
\beq
T = T^{\rm Born} + T^{\rm 1-loop} + T^{\rm c.t.} \quad ,
\eeq
where the Born terms subsume the leading electric and the subleading
magnetic couplings of the photon to the nucleon/hyperon and $\gamma^*$
denotes a real ($k^2 = 0$) or virtual ($k^2 < 0$) photon. The calculation
of the Borns term is standard, for charged kaon production the SU(3) 
generalization of the Kroll--Rudermann term gives the dominant contribution
to the electric dipole amplitude. Of particular interest is the observation
that the leading P--wave multipoles for $\Sigma^0 K^+$ production
are very sensitive
to the yet unmeasured magnetic moment of the $\Sigma^0$ because it is enhanced
by the coupling constant ratio $g_{p K \Lambda} / g_{p K {\Sigma^0}} = (D+3F)/
\sqrt{3} / (F-D) \simeq -5$. The one loop graphs are also easy to calculate.
Two remarks concerning these are in order. First, the SU(3) calculation allows
one to investigate the effect of kaon loops on the SU(2) predictions [26].
As expected, it is found that these effects are small, e.g. for neutral
pion photoproduction off protons,
\beq
\label{e0pk}
E_{0+, {\rm thr}}^K = { e F M_\pi^3 \over 96\pi^2 F_\pi^3 M_K} =
0.14 \cdot 10^{-3}/ M_{\pi^+} \,\,\,,
\eeq
which is just 1/10th of the empirical value and considerably smaller than
the pion loop contribution. The result eq.(\ref{e0pk}) is in agreement with
the famous decoupling theorem. In the chiral SU(2) limit, that is for a fixed
strange quark mass, kaon loop effects must decouple, which means that they
are suppressed by inverse powers or logs of the heavy mass, here $ M_K$. 
Eq.(\ref{e0pk}) clearly shows this behaviour.
Second, the loop graphs give rise to the imaginary part of the transistion
amplitude. Here, one encounters the standard problem of CHPT, namely that
at a given order the imaginary parts are given to much less precision than
the real ones due to the ${\cal O}(p^{2N})$ suppression for $N$--loop graphs. 
One finds that these imaginary parts come out much too big, which is caused
by the pion loops.  This can be understood by considering the recattering 
graph $\gamma p  \to \pi^+ n \to Y K^+$. By virtue of the Fermi--Watson
theorem, one finds
\beq
{\rm Im}~E_{0+} = {\rm Re}~E_{0+}^{\pi^+ n} \cdot a_{\pi K} \cdot {\rm PS}
\,\,\,,
\eeq
where PS denotes the phase space allowed for the virtual pion and $a_{\pi K}$
the $\pi K$ scattering length in the respective channel. Obviously, the
initial charged pion photoproduction process is far away from its threshold,
out of the range of applicability of CHPT. In [28,29] these multipoles were 
thus taken from the SAID data base. This is similar to the procedure adopted
in the study of double neutral photopionproduction in [26]. Clearly, this
needs refinement. At next order in the chiral expansion, one has e.g. 
additional contributions from $\pi^0$ and $\eta$ rescattering graphs. Note also
that to this order, $p^3$, the loop graphs are not finite but need standard
renormalization. This can either be done by direct Feyman diagram calculation
[28,29] or by using the general method described above (this particular 
example is worked out in detail in [10]). Finally, there are the counter 
terms.  Alltogether, there are 13 various operators with unknown
low--energy constants. One combination, $d_1+d_2$, can be fixed from
the nucleon axial radius. This also constrains the yet unmeasured
$p \to \Lambda\, K^+$ transition axial radius,
\beq
\langle r_A^2 \rangle_{p \to \Lambda \, K^+} = \frac{3
 \sqrt{2}}{D+3F} (d_1 + 3 d_2) = 0.23 \ldots 0.70 \,\, {\rm fm}^2 \,\, ,
\eeq
To fix the other LECs, $d_3 , \ldots , d_{13}$,  resonance saturation
including the baryon decuplet and the vector meson nonet was used. A detailed
account of this procedure can be found in [28].
I now summarize the results for the various
final states (photoproduction case).

\noindent ${\underline{K^0 \Sigma^+}:}$ All LECs are determined by
resonance exchange. The total cross section has been calculated for the
first 100 MeV above threshold. No  data point exists in this range so
far, but soon the new ELSA data should be available. 
The electric dipole amplitude is real at threshold, we
have $E_{0+}^{\rm thr} (K^0 \Sigma^+) = 1.07 \times 10^{-3}/M_{\pi}$.
We also have given a prediction for the recoil polarization at
$E_\gamma = 1.26\,$GeV (which is the central energy of the lowest bin
of the not yet published ELSA data). 

\noindent ${\underline{K^+ \Lambda}:}$The total cross section from
threshold up to 100 MeV above is shown in Fig.4a (left panel). The lowest
bin from ELSA [27],  $E_\gamma \in [0.96,1.01]\,$GeV,  has
$\sigma_{\rm tot}~=~(1.43 \pm 0.14)\,~\mu$b, i.e. we slightly
underestimate the total cross section. In Fig.4b (left), we show the predicted
recoil polarization $P$ at $E_\gamma = 1.21\,$ GeV (which is higher in
energy than our approach is suited for). Amazingly, the shape and
magnitude of the data [27] is well described for forward angles,
but comes out on the small side for backward angles.  
Most isobar models,
that give a descent description of the total and differential 
cross sections also at higher energies, fail to explain this angular
dependence of the recoil polarization. 

\begin{figure}[h]
\vspace{9.cm}
\includegraphics{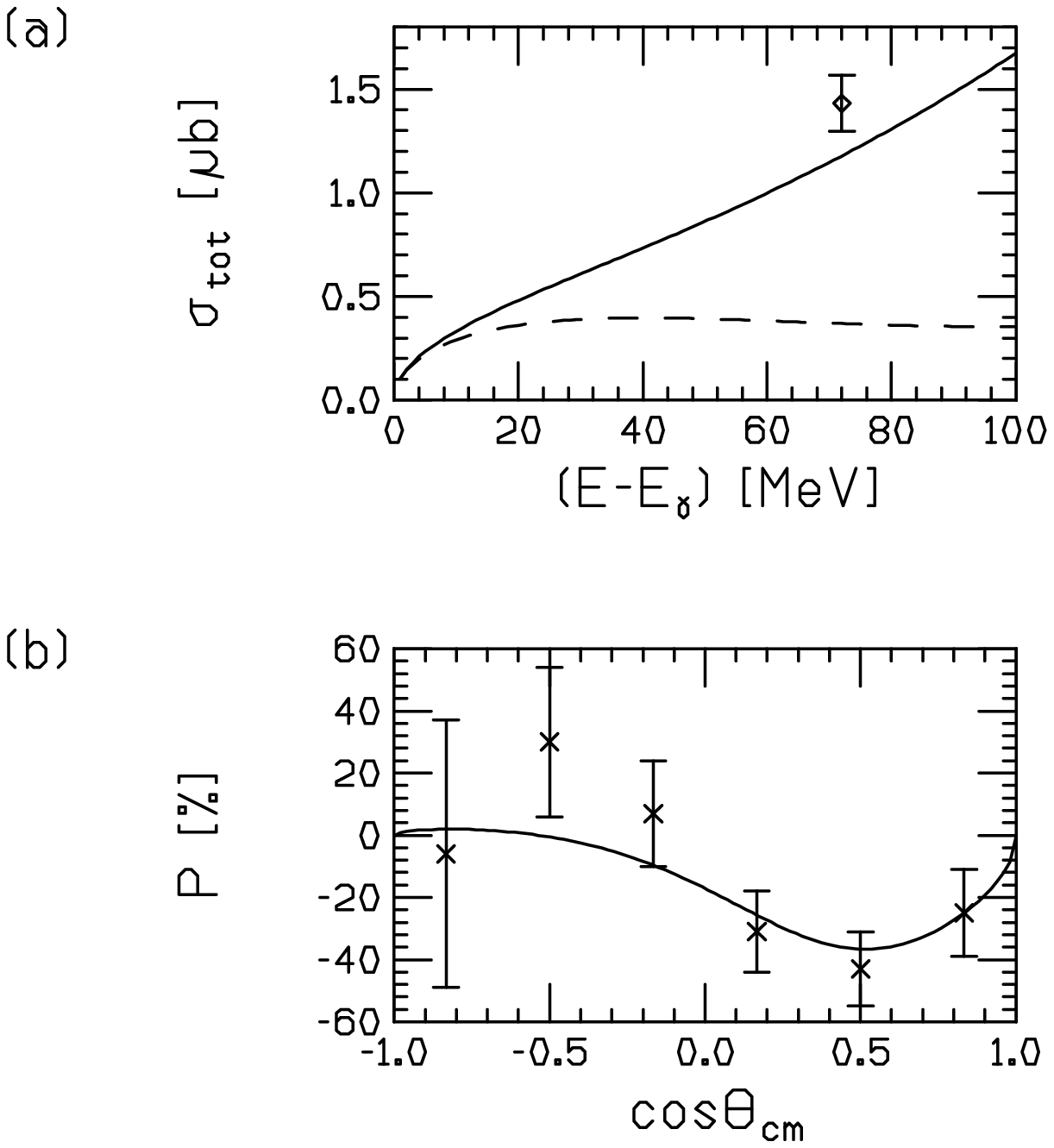}
\hspace{6.5cm}
\includegraphics{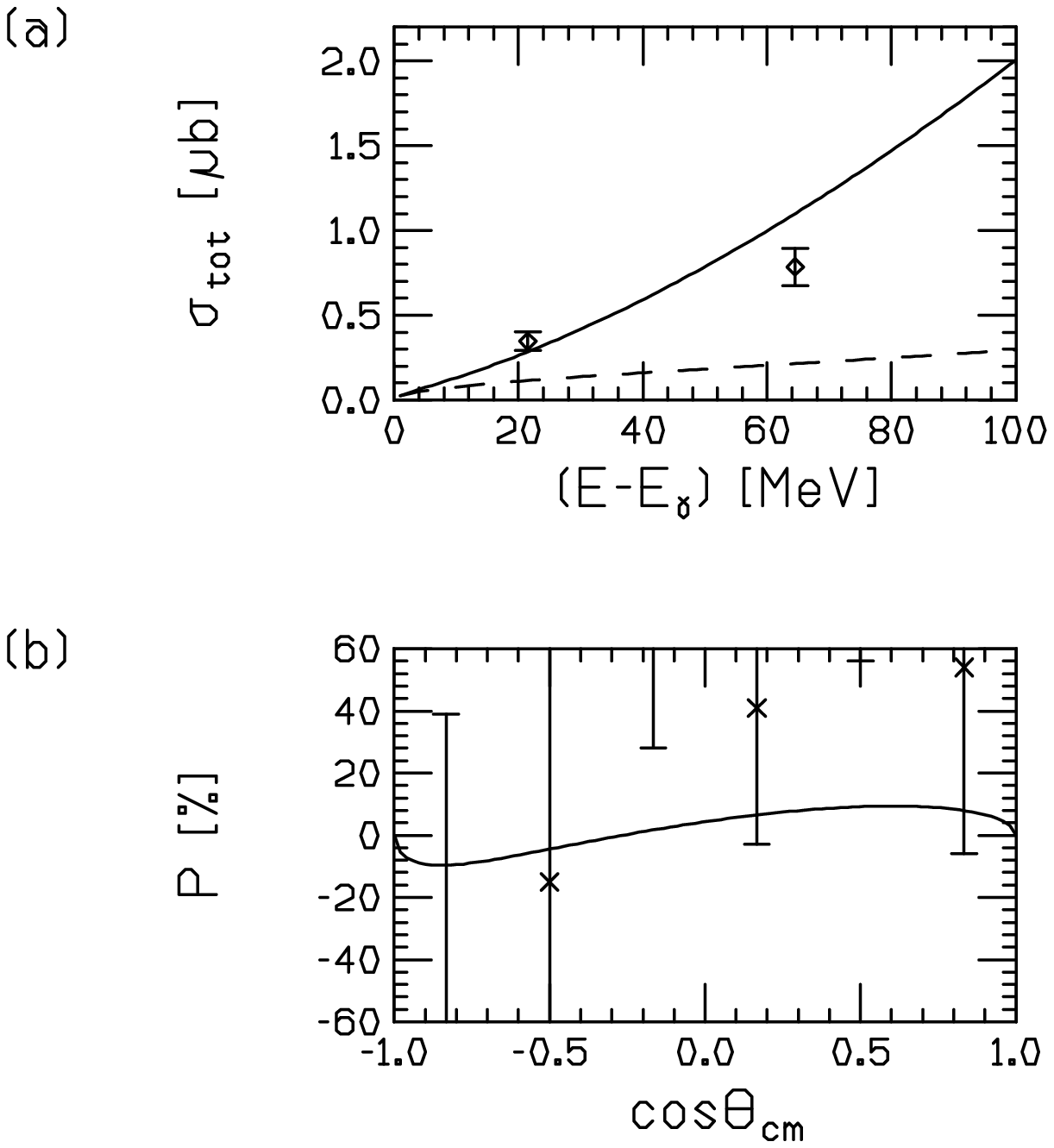}
\end{figure}

\vspace{-1.5cm}
%\caption{

\noindent Fig.~4:
Left panel: (a) Total cross section for $\gamma p \to K^+ \Lambda$
(solid line). The S--wave contribution is given by the dashed line.
(b) Recoil polarization at $E_\gamma = 1.21 \,$GeV. 
Right panel: (a) Total cross section for $\gamma p \to K^+ \Sigma^0$.
(b) Recoil polarization at $E_\gamma = 1.26 \,$GeV. 
The data are from [27].
%}

\medskip

\noindent ${\underline{K^+ \Sigma^0}:}$The total cross section is
shown in Fig.4a (right panel). It agrees  with the two data points from 
ELSA [27]. The recoil polarization at $E_\gamma =
1.26\,$GeV is shown in Fig.4b (right). It has the right shape but comes out
too small in magnitude. Nevertheless, we observe the important sign 
difference to the $K^+ \Lambda$ case, which is commonly attributed to 
the different quark spin structure of the $\Lambda$ and the 
$\Sigma^0$. Notice that this argument is strictly correct
for massless quarks only.
Here, it stems from an intricate interference of the complex S-- and
P--wave multipoles. 
%In particular, one needs both s-- and t--channel
%resonance excitations to get the shape of $P$.
In any case, one would like to have data closer to
threshold and with finer energy binning to really test the CHPT scheme.

Clearly, these results 
should only be considered indicative since we have to include (a) higher 
order effects (for both the S-- and P--waves), (b) higher partial waves
and (c) have to  get a better handle
on the ranges of the various coupling constants. In addition, one would
also need more data closer to threshold, i.e. in a region where the
method is applicable. However, the results presented are encouraging 
enough  to pursue a more detailed study of these reactions (for real 
and virtual photons) in the framework of chiral perturbation theory.

\medskip %\bigskip

\noindent 

\vspace{0.5cm}

\noindent {\bf 6 $\quad$ Two other developments}
 
\vspace{0.3cm}

\noindent Here, I will briefly draw the attention towards two other
interesting developments, namely the consistent inclusion of the
$\Delta (1232)$ resonance in the effective field theory and a coupled
channel approach to deal with the three flavor sector.

\medskip

\noindent {\it Inclusion of the $\Delta$:} Among all the resonances, 
the $\Delta (1232)$ plays a particular
role for essentially {\it two} reasons. First, the $N\Delta$ mass
splitting is a small number on the chiral scale of 1~GeV,
\begin{equation}
\Delta \equiv m_\Delta - m_N = 293 \, {\rm MeV} \simeq 3 F_\pi \,\, ,
\end{equation}
and second, the couplings of the $N\Delta$ system to pions and photons
are very strong, e.g.
\begin{equation}
g_{N\Delta \pi}  \simeq 2 g_{NN\pi} \,\, .
\end{equation}
So one could consider $\Delta$ as a small parameter. It is, however,
important to stress that in the chiral limit, $\Delta$ stays finite
(like $F_\pi$ and unlike $M_\pi$). Inclusion of the spin--3/2 fields
like the $\Delta (1232)$ is therefore based on phenomenological
grounds but also supported by large--$N_c$ arguments since in that
limit a mass degeneracy of the spin--1/2 and spin--3/2 ground state
particles appears. Recently, Hemmert, Holstein and Kambor [30]
proposed a systematic way of including the $\Delta (1232)$ based on an 
effective Lagrangian of the type ${\cal L}_{\rm eff} [U, N , \Delta ]$
which has a systematic ``small scale expansion'' in terms of {\it
  three} small parameters (collectively denoted as $\epsilon$),
\begin{equation}
\frac{E_\pi}{ \Lambda} \,\, , \quad \frac{M_\pi}{\Lambda} \,\, , \quad
\frac{ \Delta}{\Lambda} \,\, ,
\end{equation}
with $\Lambda \in [ M_\rho, m_N, 4\pi F_\pi ]$. Starting from the
relativistic pion--nucleon-$\Delta$ Lagrangian, one writes the
nucleon ($N$) and the Rarita--Schwinger  ($\Psi_\mu$)
fields  in terms of velocity eigenstates (the nucleon four--momentum
is $p_\mu = m v_\mu + l_\mu$, with $l_\mu$ a small off--shell 
momentum, $v \cdot l \ll m$ and similarly for the $\Delta (1232)$ [8]),
\beq
N = {\rm e}^{-imv \cdot x} \, (H_v + h_v) \, , \,\,\,
\Psi_\mu = {\rm e}^{-imv \cdot x} \, (T_{\mu \,v} + t_{\mu \,v}) \, ,
\eeq
and integrates out the ``small'' components $h_v$ and $ t_{\mu \,v}$
by means of the path integral formalism developed in [7]
The corresponding heavy baryon effective field theory 
in this formalism does not only
have a consistent power counting but also $1/m$ suppressed vertices
with fixed coefficients that are generated correctly (which is much simpler
than starting directly with the ``large'' components and fixing these
coefficients via reparametrization invariance). Since the spin--3/2
field is heavier than the nucleon, the residual mass difference 
$\Delta$ remains in the spin--3/2 propagator and one therefore has to
expand in powers of it to achieve a consistent chiral power counting.
The technical details how to do that, in particular how to separate
the spin--1/2 components from the spin--3/2 field, are given in
[30]. The method has been applied to the electric dipole amplitude in
threshold $\pi^0$ photoproduction [31] and other applications will
be published soon (like the $E2/M1$ ratio in the
resonance region, the corrections to the P--wave LETs in neutral
pion photoproduction and real as well as virtual Compton scattering).
 
\medskip

\noindent {\it SU(3) with coupled channels:} It is well known that in
certain reaction channels involving $K$ and $\eta$ mesons one
encounters resonances very close to the respective thresholds. The
most prominent example is the {\it sub}threshold $\Lambda (1405)$
state seen in $K^- p$ scattering.  One might therefore question the
applicability of three flavor CHPT as described before. The Munich
group has set up a scheme to deal wich such situations. It inputs the
most general
dimension two chiral meson--baryon Lagrangian into a multi--channel
S--matrix [32]. For the case of  $K^- p$ scattering one has e.g. to
take into account the pion--hyperon channels ($\pi \Lambda, \pi \Sigma$). The
couplings between the various channels are determined by the chiral
Lagrangian in terms of a few LECs. These build up the potential matrix
$V_{ij}$, which is then iterated to all orders by means of a
Lippmann--Schwinger equation,
\beq
T_{ij} = V_{ij} = \frac{2}{\pi} \sum_n \int_0^\infty dl \,
\frac{l^2}{k_n^2 -l^2} \biggl( \frac{\alpha_n^2 + k_n^2}{\alpha_n^2
  +l^2} \biggr)^2 \, V_{in} \, T_{nj} \,\, ,
\eeq
with $k_n$ the relative meson--baryon momentum in the reaction channel
$n$.
The range parameters $\alpha_n$ are necessary to render the integrals
finite, for physical reasons they are of a size of a few hundred MeV.
The resulting multi-channel S--matrix is then exactly unitary in the
subspace of the open channels. Clearly, such a unitarization procedure
induces some model--dependence and thus goes beyond the strict chiral counting.
A best fit to the data in the various
$K^- p$ reaction channels ($K^- p$, $\bar{K^0} n$, $\pi^+ \Sigma^-$,
$\pi^0 \Sigma^0$,  $\pi^- \Sigma^+$,  $\pi^0 \Lambda$) and to
threshold branching fractions allows one to fix the various LECs. It is 
interesting to note that the resulting values are comparable to the
ones found by means of resonance saturation [15]. In a similar
fashion, one can deal with pion--induced $\eta$ and $K$ meson
production as well as photopionproduction of the strange Goldstone
bosons [33]. With a few parameters, one is then able to describe a
wealth of data for a large energy range (at present, these
calculations, however, only include the S--wave amplitudes). The most
prominent result of this approach is the finding that the $\Lambda
(1405)$ as well as the $S_{11} (1535)$ should be interpreted as 
(instable) meson--baryon boundstates.   This is due to the fact that
the chiral SU(3) dynamics predicts strongly 
attractive meson--baryon interaction in the antikaon--nucleon channel
with total isospin $I = 0$ and in the kaon--$\Sigma$ channel with $I =
1/2$.

\medskip %\bigskip

\noindent 

\vspace{0.5cm}

\noindent {\bf 7 $\quad$ Short summary and outlook}

\vspace{0.3cm}

\noindent In my review talk about the status of baryon CHPT A.D. 1994 [34]
I had addressed five open problems. Referring to that paper, let me
briefly state what has happened in the mean time. (i) can  be
considered solved [35]. (ii) More precise data have been and are being
produced and as discussed, a certain understanding of the LECs in
terms of resonance exchange emerges. (iii) Similarly, first fully complete
${\cal O}(p^4)$ SU(3) calculations are becoming available supplemented
by the coupled channel approach and we will soon be able to make 
more stringent statements about the precision of the approach. (iv)
The $\Delta$ can now be handled systematically and results of many 
calculations are expected soon. Finally, (v) is still in a very infant
stage despite some impressive calculations like in [36] and some
intriguing novel approaches like e.g. in [37]. So clear
progress has been made but still lots of work remains to be done.

\vspace{1.5cm}
%\vfill

\noindent {\bf  Acknowledgements}
 
\vspace{0.3cm}

\noindent It is a pleasure to thank the organizers, in particular Prof.
Yuji Koike, for the invitation and their kind hospitality. I also would 
like to thank my collaborators V\'eronique Bernard, Bugra Borasoy,
Thomas Hemmert, Norbert Kaiser, Joachim Kambor,
Guido M\"uller and Sven Steininger for sharing with me
their insight into chiral dynamics. They can, however, not be held
responsible for any error in my presentation. I am also grateful to
Martin Moj\v zi\v s for communicating his results prior to publication.

\vspace{1.5cm}

%%references

\baselineskip 10pt

\noindent{\bf References} 

\bigskip\medskip

\noindent $\,\,\,$1. V. Bernard, N. Kaiser and Ulf-G. Mei{\ss}ner, 
{\it Int. J. Mod. Phys.\/} {\bf E4}, 193 (1995). \smallskip

\noindent  $\,\,\,$2. V. Bernard, N. Kaiser and Ulf-G. Mei{\ss}ner, preprint
KFA-IKP(TH)-1996-14 

$\, $ [hep-ph/9611253]. \smallskip

\noindent  $\,\,\,$3. V. Bernard, N. Kaiser and Ulf-G. Mei{\ss}ner, 
{\it Phys. Lett.} {\bf B309}, 421 (1993); 

$\,\,${\it Phys. Rev.} {\bf C52}, 2185 (1995). \smallskip

\noindent  $\,\,\,$4. D. Sigg et al., {\it Nucl. Phys.} {\bf A609}, 
269 (1996). \smallskip

\noindent  $\,\,\,$5. G. H\"ohler,  in  Landolt-B\"ornstein, Vol. 9b2, 
ed H. Schopper (Springer, Berlin, 1983).   \smallskip

\noindent  $\,\,\,$6. M. Moj\v zi\v s, ``Elastic $\pi$N Scattering to 
order ${\cal O}(p^3)$ in Heavy Baryon Chiral Perturbation 

$\, $  Theory'', Bratislava preprint, in preparation.
\smallskip

\noindent  $\,\,\,$7. V. Bernard, N. Kaiser, J. Kambor and Ulf-G. Mei\ss ner, 
{\it Nucl. Phys.} {\bf B388}, 315 (1992). \smallskip

\noindent  $\,\,\,$8. E. Jenkins and A.V. Manohar, Phys. Lett. 
{\bf B255}, 558 (1991).
\smallskip

\noindent  $\,\,\,$9. E. Jenkins and A.V. Manohar, in ``Effective Field
Theories of the Standard Model'',

$\,$ ed. Ulf-G. Mei{\ss}ner, World
Scientific, Singapore, 1992.\smallskip

\noindent 10. G. M\"uller and Ulf-G. Mei{\ss}ner, preprint KFA-IKP(TH)-1996-07
[hep-ph/9610275],

$\,$ accepted for publication in {\it Nucl. Phys.} {\bf B} (1997).
\smallskip

\noindent 11. G. Ecker, {\it Phys. Lett.} {\bf B336}, 508 (1994).
\smallskip

\noindent 12. G. Ecker and M. Moj\v zi\v s, {\it Phys. Lett.}
{\bf B365}, 312 (1994). \smallskip

\noindent 13. J. Gasser, {\it Ann. Phys.\,}(NY) {\bf 136}, 62 
 (1981). \smallskip

\noindent 14. P. Langacker and H. Pagels, {\it Phys. Rev.} {\bf D8}, 
4595 (1971). \smallskip

\noindent 15. B. Borasoy and Ulf-G. Mei{\ss}ner, {\it Ann. Phys.\,}(NY) 
 (1997) in press. \smallskip

\noindent 16. R. Baur and R. Urech, {\it Phys. Rev.} {\bf D53}, 6552 (1996).  
 \smallskip

\noindent 17. V. Bernard, N. Kaiser and Ulf-G. Mei\ss ner, 
{\it Z. Phys.} {\bf C60}, 111 (1993).   \smallskip

\noindent 18. E. Jenkins and A.V. Manohar, {\it Phys. Lett.} {\bf B281},
 336 (1992).  \smallskip

\noindent 19. J. Gasser, H. Leutwyler and M.E. Sainio,
{\it Phys. Lett.} {\bf B253}, 252, 260 (1991). \smallskip  

\noindent 20. V. Bernard, N. Kaiser and Ulf-G. Mei{\ss}ner, 
{\it Phys. Lett.} {\bf B389}, 144 (1996). \smallskip  
  
\noindent 21. S. Coleman and S.L. Glashow, {\it Phys. Rev. Lett.} 
{\bf 6}, 423 (1961).\smallskip

\noindent 22. D.G. Caldi and H. Pagels, {\it Phys. Rev.} {\bf D10}, 3739
(1974). \smallskip

\noindent 23. E. Jenkins, M. Luke, A.V. Manohar and M. Savage, 
{\it Phys. Lett.} {\bf B302}, 482 (1993);

$\,$ (E) {\it ibid} {\bf B388}, 866 (1996).
\smallskip

\noindent 24. Ulf-G. Mei{\ss}ner and S. Steininger, 
preprint KFA-IKP(TH)-1997-02
[hep-ph/9701260]. \smallskip

\noindent 25. D.B. Leinweber, R.M. Woloshyn and T. Draper,
{\it Phys. Rev.} {\bf D43}, 1659 (1991). \smallskip

\noindent 26. V. Bernard, N. Kaiser and Ulf-G. Mei{\ss}ner, 
{\it Phys. Lett.} {\bf B378}, 332 (1996); 

$\,\,${\it Z. Phys.} {\bf C70}, 483 (1996);
{\it Phys. Lett.} {\bf B382}, 19 (1996); {\bf B383}, 116 (1996);

$\,\,${\it Nucl. Phys.} {\bf A607}, 379 (1996). \smallskip

\noindent 27. M. Bockhorst et al., {\it Z. Phys.} {\bf C63}, 37 (1994).
 \smallskip

\noindent 28. S. Steininger, Bonn University preprint TK--96-26 (1996).
\smallskip

\noindent 29. S. Steininger and Ulf-G. Mei{\ss}ner, {\it Phys. Lett.} 
{\bf B}  (1997) in press. \smallskip

\noindent 30. T.R. Hemmert, B.R. Holstein and J. Kambor, 
"Chiral Lagrangians and $\Delta(1232)$

$\,$ interactions", in preparation.\smallskip

\noindent 31. T.R. Hemmert, B.R. Holstein and J. Kambor, 
{\it Phys. Lett.} {\bf B} (1997), in print.\smallskip

\noindent 32. N. Kaiser, P.B. Siegel  and W. Weise, 
{\it Nucl. Phys.} {\bf A594}, 325 (1995).\smallskip

\noindent 33. N. Kaiser, P.B. Siegel  and W. Weise, 
{\it Phys. Lett.} {\bf B362}, 23 (1995);

$\, $ Th. Waas, N. Kaiser and W. Weise, {\it ibid} {\bf B365}, 12 (1996);
      {\bf B379}, 34 (1996);

$\, $ {\it Nucl. Phys.} {\bf A} (1997) in print.\smallskip 

\noindent 34. Ulf-G. Mei{\ss}ner, {\it Czech. J. Phys.} 
{\bf 45}, 153  (1995). \smallskip

\noindent 35. V. Bernard, N. Kaiser and Ulf-G. Mei{\ss}ner, 
{\it Nucl. Phys.} {\bf B457}, 147 (1995). \smallskip

\noindent 36. C. Ordonez, L. Ray and U. van Kolck, {\it Phys. Rev.} 
{\bf C53}, 2086 (1996).\smallskip

\noindent 37. D. B. Kaplan, M. J. Savage and M. B. Wise,
{\it  Nucl. Phys.} {\bf B478}, 629 (1996).

\end{document}